\documentclass[a4paper,11pt]{article}
\usepackage{setspace}
\usepackage[top=2.54cm,bottom=2.54cm, left=2.2cm, right=2.2cm]{geometry}
\usepackage{bm}
\usepackage{natbib}
\usepackage{graphicx}
\usepackage{epstopdf}
\usepackage{amsmath}
\usepackage{hyperref}
\usepackage{caption}
\usepackage{amsfonts}
\usepackage{theorem} 
\theoremstyle{break}

\usepackage{mathpazo}
\usepackage{xcolor}

\begin{document}

\title{A Dirichlet Regression Model for Compositional Data with Zeros}

\author{Michail Tsagris \\ University of Crete \and Connie Stewart \\
University of New Brunswick
}

\maketitle

\begin{abstract}
Compositional data are met in many different fields, such as economics, archaeometry, ecology, geology and political sciences. Regression where the dependent variable is a composition is usually carried out via a log-ratio transformation of the composition or via the Dirichlet distribution. However, when there are zero values in the data these two ways are not readily applicable. Suggestions for this problem exist, but most of them rely on substituting the zero values. In this paper we adjust the Dirichlet distribution when covariates are present, in order to allow for zero values to be present in the data, without modifying any values. To do so, we modify the log-likelihood of the Dirichlet distribution to account for zero values. Examples and simulation studies exhibit the performance of the zero adjusted Dirichlet regression.\\
\\
\textbf{Keywords}: Compositional data, regression, Dirichlet distribution, zero values
\end{abstract}

\section{Introduction}
Compositional data are non-negative multivariate data that sum to the same constant, usually taken to be unity for convenience purposes. Their sample space is the standard simplex
\begin{eqnarray*}
\mathbb{S}^d=\left\lbrace(y_1,...,y_D)^T \bigg\vert y_i \geq 0,\sum_{i=1}^Dy_i=1\right\rbrace, 
\end{eqnarray*}
where $D$ denotes the number of variables (better known as components) and $d=D-1$. 

A natural candidate distribution for analyzing compositional data is the Dirichlet, since its support is the simplex. Dirichlet regression analysis, for the case when covariates are present, has been considered in \citet{gueorguieva2008} and \citet{hijazi2009}. A more popular type of regression was suggested by \citet{ait2003} who applied a log-ratio transformation, mapping the data from the simplex onto the Euclidean space and then applying standard multivariate regression techniques. Back transformation of the fitted values results in the fitted compositional values. 

The problem however, with both Aitchison's and Dirichlet regression, is that they are not compatible with zero values. If zero values are present in the data, the logarithm of the data is not applicable and these two regression models cannot be readily applied. For this reason, model based zero value imputation techniques have been developed \citep{palarea2008}. These methods must be applied prior to the regression analysis and assume that zeros are due to rounding or measurement errors. Rather than making an assumption concerning whether they are rounded or not, here we propose adjusting the Dirichlet log-likelihood to accommodate the zero values. 

Dirichlet regression when zeros are not present using a parametrization of the Dirichlet distribution is reviewed in Section 2. The general problem of zeros in the analysis of compositional data is briefly discussed in Section 3. In the same Section the adjustment of the log-likelihood of the Dirichlet regression model needed to handle zeros is described and two final models are presented. In Section 4 we compare the zero adjusted Dirichlet regression models with two other models, using real-life data examples as well as small scale simulation studies to assess the performance of the Dirichlet models. A discussion of the advantages and limitations of our methodology in the context of related work is presented in Section 5, as well as some research directions for compositional data regression. Section 6 closes the paper with some final conclusions. 

\section{Dirichlet regression} 
A composition $\mathbf{Y}$ is Dirichlet distributed (that is, $\mathbf{Y}\sim \mathrm{Dir}(\phi,\mathbf{a}^*)$) if its density can be written as
\begin{eqnarray*}
f\left({\bf y} \right)=\frac{\Gamma\left(\sum_{i=1}^D\phi a^*_i\right)}{\prod_{i=1}^D\Gamma\left(\phi a^*_i\right)}\prod_{i=1}^Dy_i^{\phi a^*_i-1},
\end{eqnarray*}
and $\sum_{i=1}^Da^*_i=1$. We first consider the case where only $\mathbf{a^*}$ depends on a set of predictor variables, and subsequently the case where both $\phi$ and $\mathbf{a}^*$ are linked to covariates. In order to relate $\mathbf{a^*}$ to a set of $p$ independent variables ${\bf X}=\left(1,X_1,\ldots,X_p\right)$, we use the following link function
\begin{eqnarray} 
\begin{array}{ll} \label{regalpha}
a^*_1 =\frac{1}{1+\sum_{k=2}^D e^{{\bf x}^T\pmb{\beta}_k}} & \\
a^*_i = \frac{e^{{\bf x}^T\pmb{\beta}_i}}{1+\sum_{k=2}^De^{{\bf x}^T\pmb{\beta}_k}},\  \text{for\ } i=2,...,D,
\end{array}
\end{eqnarray}
where 
\[\pmb{\beta}_i=\left(\beta_{0i},\beta_{1i},...,\beta_{pi} \right)^T, \ i=2,...,D.\]
The corresponding log-likelihood is
\begin{eqnarray} \label{dirireg2}
\ell &=& n\log{\Gamma\left(\phi\right)}-\sum_{j=1}^n\sum_{i=1}^D\log{\Gamma\left(\phi a^*_{ij}\right)}
+\sum_{j=1}^n\sum_{i=1}^D\left(\phi a^*_{ij}-1\right)\log{y_{ij}}.
\end{eqnarray}

The interpretation of the resulting regression parameters is not difficult, as will be seen later, and is the same as those of the standard linear regression model suggested by \citet{ait2003} since the same link function is used. In addition, this formulation requires $d(p+1)+1$ parameters, whereas the classical formulation of the Dirichlet requires $D(p+1)$ parameters \citep{hijazi2009, gueorguieva2008}, where $p$ denotes the number of independent variables and $d=D-1$. 

The parameter $\phi$ can also be linked to the same covariates. Instead of having the same value of $\phi$ for all compositional vectors, one can allow for it to vary as a function of the covariates. The link function used here is the logarithm to ensure that it is always positive and is given by
\begin{eqnarray} \label{phi}
\phi_j^*=e^{{\bf x}_j^T\pmb{\gamma}}
\end{eqnarray}
where  
$\pmb{\gamma}=(\gamma_0,\gamma_1,\ldots,\gamma_p)$.  A slightly more involved log-likelihood function can then be obtained by substituting the precision parameter $\phi$ in (\ref{dirireg2}) with (\ref{phi}) yielding
\begin{eqnarray} \label{dirireg3}
\ell &=& \sum_{j=1}^n\log{\Gamma\left(\phi_j^*\right)}-\sum_{j=1}^n\sum_{i=1}^D\log{\Gamma\left(\phi_j^* a^*_{ij}\right)} +
\sum_{j=1}^n\sum_{i=1}^D\left(\phi_j^* a^*_{ij}-1\right)\log{y_{ij}}.
\end{eqnarray} 
Note that this model (\ref{dirireg3}) has $p$ additional parameters compared to the classical Dirichlet regression model.

\section{Zero adjusted Dirichlet regression}
In this Section we will show how can one perform Dirichlet regression when zero values of any type are present in the data. We begin with a brief discussion on the problem of zero values in compositional data followed by the specific case of Dirichlet regression when the response is compositional with zero values.

\subsection{Zero values and current approaches}
When \citet{ait1982} suggested the log-ratio analysis of compositional data, he noted that zero values are not allowed. For this reason, he suggested  ad-hoc zero replacement strategies \citep{ait2003}. Some of them have nice properties and are preferred. For instance, the fact that logratio based Euclidean distances between two compositional data are preserved is an attractive property.

\citet{palarea2008} developed a model-based imputation of zero values and \citet{martin2012} suggested a robust version of it. The underlying assumption behind all of these strategies is that the zero values are actually unobserved very small quantities which were rounded to zero. An example is when the detection limit of a measurement instrument is not low enough. 

A parametric zero value imputation or parametric replacement of rounded zeros, using the EM algorithm, can be found in the R package \textit{robCompositions} \citep{templ2011}. After the zero value imputation or replacement, the additive log-ratio transformation (\ref{alr}) is applied to the data and a multivariate linear regression model is carried out. The fitted values are then back transformed into the simplex. 

The drawback of these techniques, however, is that all components will have to change slightly if there is at least one zero value in the compositional vector. When there are many zeros in a dataset, many other observed values will have to change as well. From one point of view this could be necessary, since the observed proportions are not the true ones. On the other hand, imputation of zero values induces some extra variability to the data. Additionally, the results of a zero value imputation (assuming the values were truncated to zero) can be sensitive to the assumed detection limit \citep{scealy2011}. In some cases it is known that the zero is truly zero (a structural zero) and changing the value is clearly not ideal. Finally, these approaches cannot handle satisfactorily a large point mass at 0.

A second direction of approaches adopted by \citet{zadora2010}, \citet{scealy2011}, \citet{stewart2011}, and most recently by \citet{bear2015}, (Com 20) is to explicitly incorporate the zero values, without having to impute them.For instance, \citet{zadora2010} and \citet{bear2015} model the probability of a zero value separately; a method suggested by \citet[~p. 271-273]{ait1982} as well. 
The square root transformation in \citet{scealy2011} maps the zero values on the surface of a hyper-sphere, thus they are treated as allowable points on the hyper-sphere. \citet{scealy2011} performed regression by employing the Kent distribution and their approach falls within the category of treating compositional data as directional data through the square root transformation (see also \citet{stephens1982}). Other regression techniques on the simplex addressing the problem of zero values can be found in \citet{leininger2013} who implemented spatial regression for compositional data with many zeros employing a scaling power transformation and assuming a latent multivariate normal model. 

Our method, presented in detail in the following Subsection, is an extension of the basic ideas of \citet{zadora2010}, \citet{stewart2011} and \citet{bear2015} to the Dirichlet regression context. However, rather than using a log ratio approach as these authors have done, we instead utilize the Dirichlet distribution and also allow covariates to be present. Unlike \citet{scealy2011} our model is easier to fit and interpret with fewer parameters to estimate.

\subsection{Zeros values and Dirichlet regression} 

Our approach for managing the zeros relies on the methodology developed in \citet{stewart2011}, but is extended to include the Dirichlet regression setting and modified to handle some practical issues that arise when attempting to estimate the parameters of interest.   

Following \citet{stewart2011}, we assume that there are $B$ populations (or groups) corresponding to each possible subset of non-zero components of the compositional vector $\mathbf{Y}$. To set the notation, let $\mathbf{G}$  denote the vector indexing the non-zero components of $\mathbf{Y}$. Further let $\theta_b = P\left [ \mathbf{G}= \mathbf{g}_b \right ]$ (the marginal probability that an observation comes from population $b$) where $\mathbf{g}_b$ is the vector with 1s and 0s corresponding to population $b$ and  $\sum_{b=1}^B \theta_b = 1$.  The density of $\mathbf{Y}$ with non-zero components corresponding to population $b$ is then
\begin{eqnarray}
\label{eqn:joint}
f_{\mathbf{Y}}(\mathbf{y}) & = &  \sum_{b=1}^B f_{\mathbf{Y},\mathbf{G}}(\mathbf{y},\mathbf{g}_b) \\ \nonumber
						   & = & f_{\mathbf{Y},\mathbf{G}}(\mathbf{y},\mathbf{g}_{b^*})
\end{eqnarray}
where $\mathbf{g}_{b^*}$ is the vector of indices corresponding to the nonzero components of $\mathbf{y}$. Note that for $b\neq b^*$, $f_{\mathbf{Y},\mathbf{G}}(\mathbf{y},\mathbf{g}_b) = 0$.  Now let $\mathbf{y}_{b^*}$ of length $D_{b^*}$ denote the vector containing the non-zero components of $\mathbf{y}$ and $f_{b^*}(\mathbf{y}_{b^*})$ the density of $\mathbf{Y}_{b^*}$.  From Equation \ref{eqn:joint},
\begin{eqnarray}
f_{\mathbf{Y}}(\mathbf{y}) 	& = & f_{\mathbf{Y},\mathbf{G}}(\mathbf{y},\mathbf{g}_{b^*}) \\ \nonumber
							& = & f_{\mathbf{Y}|\mathbf{G}}(\mathbf{y}|\mathbf{g}_{b^*})\theta_{b^*} \\ \nonumber
                            & = & f_{b^*}(\mathbf{y}_{b^*}) \theta_{b^*} \nonumber
\end{eqnarray}

While in \citet{stewart2011} $f_{b^*}(\mathbf{y}_{b^*})$ was chosen to be the multiplicative logistic normal distribution, here we consider using the Dirichlet distribution, that is $\mathbf{Y}_{b^*} \sim \mathrm{Dir}(\phi_{b^*},\mathbf{a}^*_{b^*})$.  With this model, the density function of $\mathbf{Y}_{b^*}$ 
is then
\[ f_{b^*} (\mathbf{y}_{b^*})=\frac{\Gamma \left ( \sum_{i=1}^{D_{b^*}} \phi_{b^*} a_{b^*i}^{*} \right )}{\prod_{i=1}^{D_{b^*}}\Gamma \left (\phi_{b^*} a_{b^*i}^{*} \right )} \prod_{i=1}^{D_{b^*}}y_{b^*i}^{ \phi_{b^*} a_{b^*i}^{*}-1},\]
where, for a given set of predictor variables, $ a_{bi}^{*}$ are defined as in (\ref{regalpha}). Note that with this model, a distinct vector of regression coefficients is allowed for each population.

For a sample of compositional data, $\mathbf{Y}_1,\mathbf{Y}_2,\ldots,\mathbf{Y}_n$, and associated covariates, let $S_b$ denote the set of observations with zeros indexed by $\mathbf{g}_b$. The Dirichlet log-likelihood function in (\ref{dirireg2}), adjusted to accommodate zero values, becomes
\begin{eqnarray}  \nonumber
\label{mixdirireg2}
l &=& n_1 \mathrm{log}(\theta_1) + n_1\log{\Gamma\left(\phi_1\right)}-\sum_{\{j:\mathbf{y}_j \in\ S_1\}}\sum_{i=1}^{D_1}\log{\Gamma\left(\phi_1 a^*_{1ij}\right)}
+\sum_{\{j:\mathbf{y}_j \in\ S_1\}}\sum_{i=1}^{D_1}\left(\phi_1 a^*_{1ij}-1\right)\log{y_{1ij}} +\ldots+ \\ 
& & 
n_B \mathrm{log}(\theta_B) + n_B\log{\Gamma\left(\phi_B\right)}-\sum_{\{j:\mathbf{y}_j \in\ S_B\}}\sum_{i=1}^{D_B}\log{\Gamma\left(\phi_B a^*_{Bij}\right)}
+\sum_{\{j:\mathbf{y}_j \in\ S_B\}}\sum_{i=1}^{D_B}\left(\phi_B a^*_{Bij}-1\right)\log{y_{bij}} 
\end{eqnarray}
where for $b=1,\ldots,B$, $D_b$ denotes the number of non-zero components in population $b$, $n_b$  denotes the  number of observations in the sample from population $b$ and $a_{bi}^{*}$ is defined as in (\ref{regalpha}).  Further, $\sum_{b=1}^B n_b = n$ and $\sum_{i=1}^B \theta_b = 1$.

While the technical aspects related to maximizing the above log-likelihood are discussed in Subsection \ref{sub:max}, it is straightforward to show that $\hat{\theta}_b=n_b/n$ is the maximum likelihood estimator of $\theta_b$ for $b=1,\ldots,B$.  Furthermore, by noting that the log-likelihood is comprised of a sum of individual simple Dirichlet log-likelihoods (refer to Equation \ref{dirireg2}), each with its own set of parameters, we may at least conceptually estimate the parameters (other than $\theta_b,\ b=1,\ldots,B)$ through a Dirichlet regression analysis of the non zero components for each group.   The difficulty in practice, however, is that there may be many groups with few observations, particularly when the dimension of $\mathbf{Y}$ is large.  In this case, parameter estimates are either not able to be obtained by this approach or may be unreliable.

To deal with this issue we make use of some of the ideas proposed by \citet{bear2015} for handling zeros which, in our setting, essentially translates to the assumption that the regression parameters are the same across groups. Note that \citet{bear2015} did not consider the regression setting nor the Dirichlet distribution. Instead they proposed an extension of the additive logistic normal distribution which involves the additive logratio transformation (see \cite{ait2003}) but allows for essential zeros.  While their technique requires that the data be such that one of the components of the composition is never zero, this assumption is not needed with the Dirichlet model. 

As in \citet{bear2015}, we define a selection matrix, ${\bf Q}_b,\ b=1,\ldots,B$, which  picks out the non-zero elements of the composition ${\bf Y}$ corresponding to group $b$.  As an example with $D=4$, suppose that group one ($b=1$) consists of compositions with zeros in the first and third positions (and non-zero elements in the second and forth positions).  In this case 
\[ {\bf Q}_1= \left [ 
\begin{array}{llll} 
 0 & 1 & 0 & 0 \\
 0 & 0 & 0 & 1 
\end{array} 
\right ]
.\]

A modified log-likelihood may be obtained by considering the density of $\mathbf{Y}$ with non-zero components corresponding to population $b$ as
\[f_{\mathbf{Y}}(\mathbf{y}) = f_b(\mathbf{y}_b)\theta_{b},\]
but now with the simplifying assumption that $\mathbf{Y}_b \sim \mathrm{Dir}(\phi,{\bf Q}_b\mathbf{a}^*)$, where ${\bf Q}_b\mathbf{a}^*$ is a vector of length $D_b$ and $\mathbf{a}^*$ is a common (across groups) vector of parameters linked to covariates according to (\ref{regalpha}). Letting $\mathbf{Q}_b\mathbf{a}^*[i]$ denote the $i$th element of this vector, the modified log-likelihood becomes
\begin{eqnarray} \label{zadr} 
l &=& n\log{\Gamma\left(\phi\right)}+ n_1 \mathrm{log}(\theta_1) -
\sum_{\{j:\mathbf{y}_j \in\ S_1\}}\sum_{i=1}^{D_1}\log{\Gamma\left( \phi  \mathbf{Q}_1\mathbf{a}_j^*[i] \right)}+\sum_{\{j:\mathbf{y}_j \in\ S_1\}}\sum_{i=1}^{D_1}\left(\phi\mathbf{Q}_1\mathbf{a}_j^*[i]-1\right)\log{y_{1ij}} +\ldots \nonumber \\
& & 
+n_B \mathrm{log}(\theta_B) -\sum_{\{j:\mathbf{y}_j \in\ S_B\}}\sum_{i=1}^{D_B}\log{\Gamma\left(\phi \mathbf{Q}_B\mathbf{a}_j^*[i]\right)}
+\sum_{\{j:\mathbf{y}_j \in\ S_B\}}\sum_{i=1}^{D_B}\left(\phi \mathbf{Q}_B\mathbf{a}_j^*[i]-1\right)\log{y_{bij}}.
\end{eqnarray}
We will refer to the  Dirichlet regression based on the above model as the zero adjusted Dirichlet regression (ZADR) model. 

The analogous but more complex log-likelihood based on model (\ref{dirireg3}) is then
\begin{eqnarray} 
\label{mixdirireg3} 
l &=& \sum_{j=1}^n\log{\Gamma\left(\phi_j\right)} +  \nonumber \\
 & & n_1 \mathrm{log}(\theta_1)  -
\sum_{\{j:\mathbf{y}_j \in\ S_1\}}\sum_{i=1}^{D_1}\log{\Gamma\left( \phi_j  \mathbf{Q}_1\mathbf{a}_j^*[i] \right)} \nonumber 
 + \sum_{\{j:\mathbf{y}_j \in\ S_1\}}\sum_{i=1}^{D_1}\left(\phi_j\mathbf{Q}_1\mathbf{a}_j^*[i]-1\right)\log{y_{1ij}} + \ldots + \nonumber \\
 & & n_B \mathrm{log}(\theta_B) -\sum_{\{j:\mathbf{y}_j \in\ S_B\}}\sum_{i=1}^{D_B}\log{\Gamma\left(\phi_j \mathbf{Q}_B\mathbf{a}_j^*[i]\right)}
+\sum_{\{j:\mathbf{y}_j \in\ S_B\}}\sum_{i=1}^{D_B}\left(\phi_j \mathbf{Q}_B\mathbf{a}_j^*[i]-1\right)\log{y_{bij}},
\end{eqnarray}
where $\phi_j^*$ is given in Equation (\ref{phi}) and $\mathbf{a}^*$ is as before.  To distinguish the two models we refer to (\ref{zadr}) as simple ZADR and (\ref{mixdirireg3}) as ZADR.  As in the non zero case, the log-likelihood ratio test can be used to decide which model, the simple ZADR or ZADR, is suitable for the data. 

The fitted model can be used to estimate the compositional values when new values from the predictor variables are available. Like any other suggested model in the literature, a zero value will be difficult to be predicted. 

In the next Section we describe a way of of maximising the log-likelihood in (\ref{zadr}). The maximization of (\ref{mixdirireg3}) is almost the same.

\subsection{Maximisation of the log-likelihood}
\label{sub:max}
When a maximisation is performed numerically, there is the question of sensitivity to the initial values. We have seen empirically that this is not the case. At first we must use the zero free compositional vectors to get estimates which will be termed initial estimated coefficients. We then add the contribution to the log-likelihood of the compositional vectors having zero values in at least one of their components. Thus, the log-likelihood in (\ref{zadr}) is actually the sum of two log-likelihoods, the one due to zero free compositional vectors and the other due to the non-zero free compositional vectors. The algorithm is as follows:
\begin{enumerate}
\item[Step 1] Use the zero free compositional vectors and apply the additive log-ratio transformation \citep{ait2003}
\begin{eqnarray} \label{alr}
z_i=\log{\frac{y_i}{y_1}}, \ \ \text{where} \ \ i=2,\ldots,D
\end{eqnarray}
\item[Step 2] Estimate the parameters using least squares 
\begin{eqnarray}\label{ait}
{\bf B}=\left({\bf X}^T{\bf X}\right)^{-1}{\bf X}^T{\bf Z},
\end{eqnarray}  
where ${\bf Z}$ is the $n \times d$ matrix containing the log-ratio transformed composition and ${\bf X}$ is the $n \times p$ design matrix. The fitted parameters ${\bf B}$ in (\ref{ait}) result in the Aitchison model. 
\item[Step 3] Use the estimates from Step 2 as initial values for maximising the log-likelihood of the Dirichlet regression in (\ref{dirireg2}). The estimates are the initial estimated coefficients. If the  model in (\ref{dirireg3}) is applied, the initial values for the precision parameter $\phi$ are randomly chosen.
\item[Step 4] Use the estimated coefficients from Step 3 as initial values in maximising the ZADR model (\ref{zadr}), and obtain the final estimated coefficients. 
\end{enumerate}

The advantage of using the above described procedure is that we have good starting values and reduce the probability of obtaining a local instead of a global maximum of the relevant log-likelihood.

\section{Real data analysis and simulation studies}

In order to assess the usefulness of the ZADR models, we consider four real-life data sets followed by a small scale simulation study in which we compare these models to other approaches (such as Aitchison's log-ratio method with zeros modified and the Kent model used in \citet{scealy2011}). For each model, we measure its performance by  computing the Kulback-Leibler divergence of the observed data from the fitted data, as well as the $L_2$ norm.  These are given by 
\begin{eqnarray} \label{KL}
KL\left({\bf y},\hat{{\bf y}}\right)=\sum_{i=1}^n\sum_{j=1}^Dy_{ij}\log{\frac{y_{ij}}{\hat{y}_{ij}}} \ \ \text{and} \ \ 
\|{\bf y}-\hat{{\bf y}}\|_2=\sum_{i=1}^n\left({\bf y}_i-\hat{{\bf y}}_i\right)^T\left({\bf y}_i-\hat{{\bf y}}_i\right) \ \ \text{respectively}.
\end{eqnarray}

Note that the Kullback-Leibler divergence is defined even when $y_{ij}=0$, simply because $0\log(0)=0$. The contribution of $y_{ij}$ is $0$ in this case. Since the Aitchison model requires that all values be strictly positive, throughout all the forthcoming examples, zero values will be imputed prior to application of this model. The R package \textit{robCompositions} \citep{templ2011} will be used for this purpose.
For all models in the test bed, the Kulback-Leibler divergence and the $L_2$ norm were calculated using the initial compositional data, i.e. with no zero value imputations.

\subsection{Real data analysis 1: Presidential election data}
Using data from a political science application, we will compare the simple ZADR model against Aitchison's model. The results of the 2000 U.S. presidential election results in the 67 counties of Florida were analysed by \citet{smith2002}. The total number of votes for each of the 10 candidates is available, along with some demo-graphical information (8 variables) for each county. The data are accessible at \href{http://www.stat.unc.edu/faculty/rs/palmbeach.html}{http://www.stat.unc.edu/faculty/rs/palmbeach.html}. 

There were 23 counties with at least one zero value, meaning that in some counties, mainly one or two candidates received zero votes. In relative figures, there were about 5\% zero values. For each county we divided the total votes of each candidate by the total of the votes in order to turn them into compositions. 

The Kullback-Leibler divergence for the simple ZADR and Aitchison's model were 0.844 and 0.648 respectively. As for the $L_2$ norm, the corresponding relevant numbers were 0.519 and 0.513.

It is interesting to note that if we convert the fitted values into estimated frequencies, by multiplying the estimated composition in each county by the number of votes, and calculate the $\chi^2$ distance between the observed and the estimated number of votes, the simple ZADR model seems to do slightly better.   

We also fitted a Dirichlet-multinomial distribution to the original data and the Kullback-Leibler divergence and $L_2$ norm were 0.638 and 0.512, respectively, indicating better fits than the simple ZADR and Aitchison models. This example shows clearly that multinomial, Dirichlet-multinomial or in general frequency related data can also be analysed with compositional data developed regression models and zeros can also be handled. 

\subsection{Real data analysis 2: Glass data} 
In this second example we compare the simple ZADR model and Aitchison's approach using the forensic glass dataset which has $n=214$ observations from $6$ different categories of glass, where each observation is a composition with $D=8$ components. In total there are $392$ (22.90\%) zero values and Table \ref{zeros2} shows in which components these zeros arise. The data are available from \href{https://archive.ics.uci.edu/ml/datasets/Glass+Identification}{https://archive.ics.uci.edu/ml/datasets/Glass+Identification}.

\begin{table}[ht]
\begin{center}
\begin{tabular}{c|c|c|c|c} \hline \hline
Components           & Sodium    & Magnesium & Aluminium & Silicon  \\ \hline    
Percentage of zeros  & 0.00\%    & 19.63\%   & 0.00\%    & 0.00\%    \\ [5pt] 
Components           & Potassium & Calcium   & Barium    & Iron       \\ \hline 
Percentage of zeros  & 14.02\%   & 0.00\%    & 82.24\%   & 67.29\%    \\ \hline \hline
\end{tabular}       
\caption{Forensic glass data: the percentage of observations for which each component is zero.}
\label{zeros2}
\end{center}
\end{table}

The only covariate we could use was the the refractive index of each glass (the glass categories made the design matrix of the non zero compositions ill-posed). The Kullback-Leibler divergence for the simple ZADR and the Aitchison model were 4.02 and 3.95 respectively. The $L_2$ norms were 0.142 and 0.137 for the simple ZADR and the Aitchison model respectively. Table \ref{glass} shows the parameters of the simlple ZADR and the Aitchison regression model.

\begin{table}[ht]
\begin{center}
\begin{small}
\begin{tabular}{l|c|c|c|c|c|c|c} \hline \hline
\multicolumn{8}{c}{\textbf{Simple ZADR model}} \\ [5pt]
Response  & \multicolumn{7}{c}{$\log{(Element/Sodium)}$}   \\ \hline \hline
Element   &  Magnesium  &  Aluminium  & Silicon  & Potassium  &  Calcium &  Barium  &  Iron \\ \hline  
Constant  & -1.394(0.018) & -2.223(0.023) & 1.686(0.008) & -3.251(0.038) & -0.425(0.012) & -2.908(0.070) & -4.141(0.088)  \\ [5pt]
Slope     &  -0.015(0.007) & -0.046(0.008) & 0.002(0.003) & -0.123(0.016) & 0.042(0.004) & 0.076(0.024) & 0.020(0.023)  \\ [5pt] 
\multicolumn{8}{c}{\textbf{Aitchison model}} \\ [5pt]
Response  & \multicolumn{7}{c}{$\log{(Element/Sodium)}$}   \\ \hline \hline
Element   &  Magnesium  &  Aluminium  & Silicon  & Potassium  &  Calcium &  Barium  &  Iron \\ \hline  
Constant  & -1.947(0.081) & -2.271(0.024) & 1.691(0.004) & -3.944(0.105) & -0.428(0.008) & -5.906(0.101) & -7.542(0.152) \\ [5pt]
Slope     & -0.060(0.026) & -0.046(0.008) & 0.002(0.001) & -0.080(0.034) & 0.042(0.003) & -0.118(0.033) & 0.165(0.05)   \\ \hline \hline
\end{tabular}       
\caption{Forensic glass data: estimated regression parameters along with their standard errors inside the parentheses.}
\label{glass}
\end{small}
\end{center}
\end{table}
The Wald test for each of the coefficients show that all of them are significant for the Aitchison model apart from the Silicon element. The same is true for the simple ZADR model which also identifies the Iron element as non significant. 

\subsection{Real data analysis 3: Foraminiferal data}
This third dataset consists of foraminiferal (marine plankton species) compositions measured at $30$ different depths ($1$-$30$ metres) and can be found in \citet{ait2003}. The data consist of 4 components, Triloba, Obesa, Pachyderma and Atlantica. There are $5$ compositional vectors having a zero value, either in the third or the fourth component. The data were also analysed by \citet{scealy2011} using the logarithm of the depth as the independent variable. Since there are $4$ components we cannot plot them on a triangle. While we could plot them in a $3D$ plot,  when printed it would not reveal the structure of the data. Figure \ref{fig2} presents the data as a function of the water depth. It can be seen that the logarithm of the water depth does not really affect the composition. 

We first compare the analysis of the data via the simple ZADR model in (\ref{zadr}) versus the ZADR model in (\ref{mixdirireg3}), and subsequently the simple ZADR versus Aitchison's model as well as the analysis by \citet{scealy2011}. 

\subsubsection{Comparison of the ZADR Models}

\begin{table}[!ht]
\caption{ \label{tab1} Parameter estimates for the simple ZADR model applied to the foraminiferal data. The \textit{Bias} columns refer to the estimated bias of the parameters calculated using parametric bootstrap. The standard errors appear in the parentheses next to the estimates.}
\centering
\begin{tabular}{c|ccccc} \hline\hline
Response variable  & Constant       & Bias          & Slope         & Bias           \\       \hline                    
Obesa              & -1.225(0.348)  & 0.019(0.337)  & 0.117(0.131)  & -0.007(0.127)  \\ [5pt]
Pachyderma 		   & -2.392(0.463)	& 0.010(0.489)  & 0.087(0.173)  & -0.008(0.187)  \\ [5pt] 
Atlantica          & -2.298(0.494)  & 0.007(0.479)  & -0.046(0.193) & -0.009(0.185)  \\ [5pt]
$\phi$             & 15.889(2.473)  & 1.683(2.981)  &               &                \\   \hline \hline
\end{tabular}
\end{table}

\begin{table}[!ht]
\caption{ \label{tab1a} Parameter estimates for the ZADR model  applied to the foraminiferal data. The \textit{Bias} columns refer to the estimated bias of the parameters calculated using parametric bootstrap. The standard errors appear in the parentheses next to the estimates.}
\centering
\begin{tabular}{c|ccccc} \hline\hline
Response variable  & Constant       & Bias           & Slope         & Bias          \\       \hline                    
Obesa              & -1.190(0.254)  & 0.000(0.262)   & 0.101(0.104)  & 0.000(0.107)  \\ [5pt]
Pachyderma 		   & -2.744(0.428)	& -0.049(0.458)  & 0.225(0.169)  & 0.015(0.179)  \\ [5pt] 
Atlantica          & -2.563(0.430)  & -0.062(0.441)  & 0.059(0.178)  & 0.016(0.181)  \\ [5pt]
$\phi$             & 3.725(0.481)   & 0.357(0.584)   & -0.371(0.184) & -0.101(0.221)  \\   \hline \hline
\end{tabular}
\end{table}

\begin{figure}[!ht]
\centering
\begin{tabular}{cc}
\includegraphics[scale=0.5,trim=0 20 20 20]{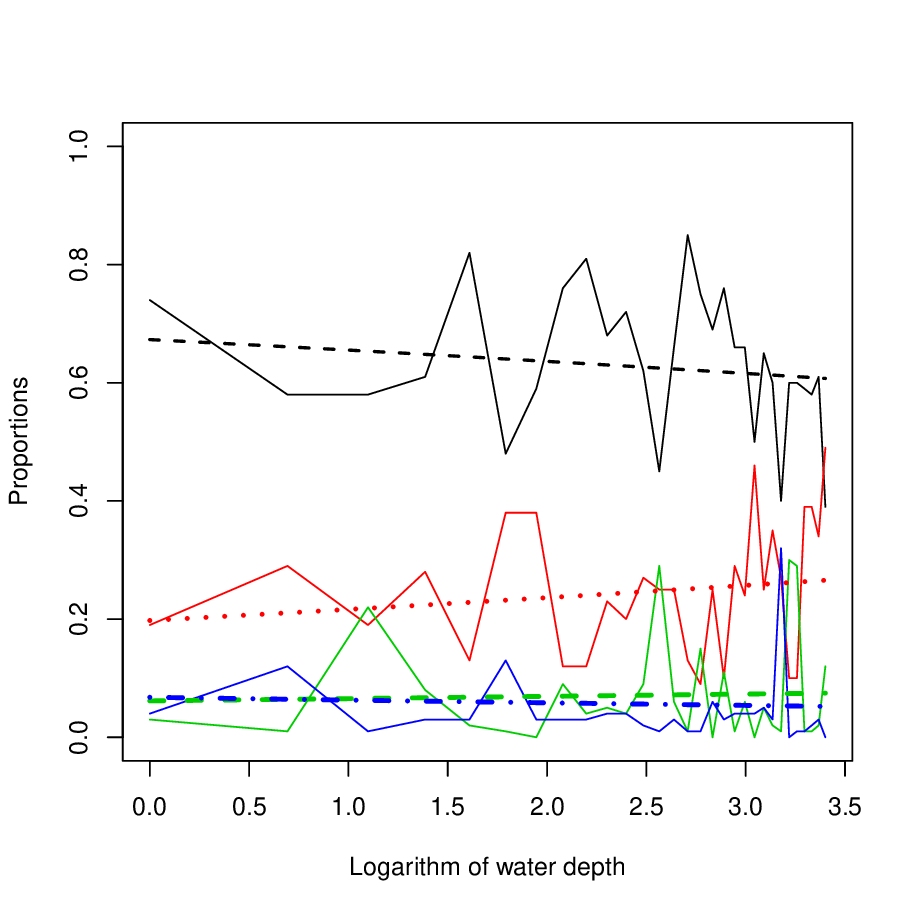} &
\includegraphics[scale=0.5,trim=0 20 20 20]{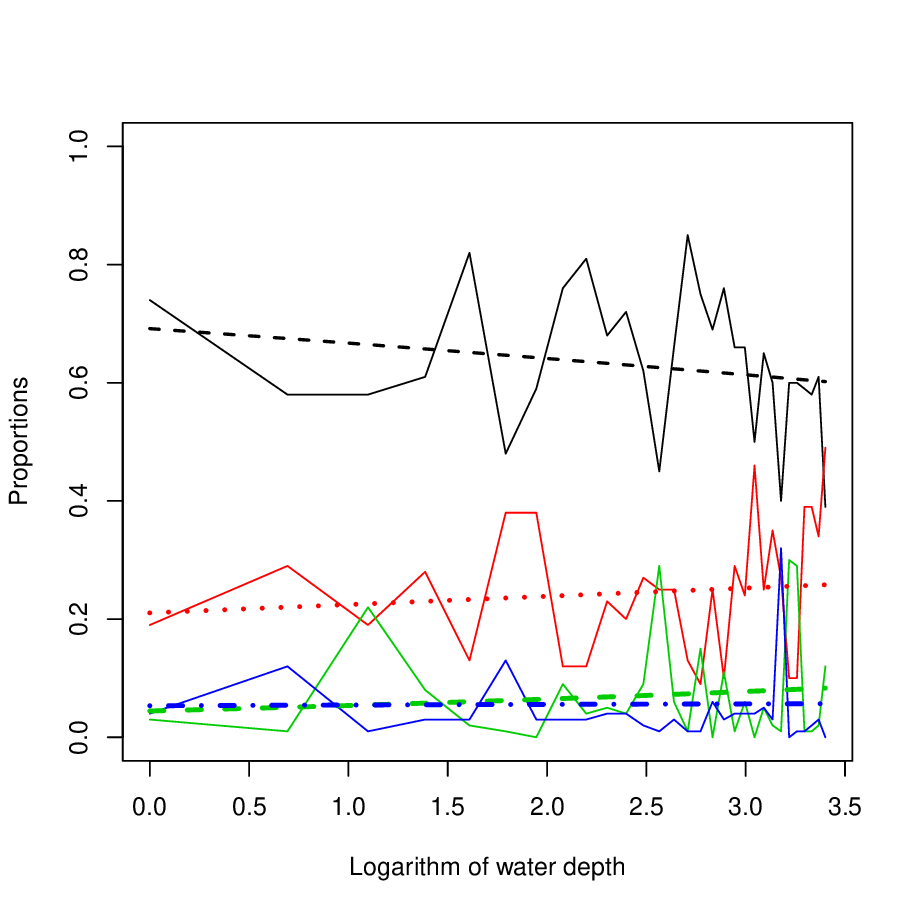}   \\
\footnotesize{(a)}   &  \footnotesize{(b)}   
\end{tabular}
\caption{Both plots show the foraminiferal compositions as a function of the water depth. As we move from left to right, we see that the logarithm of the water depth increases. The fitted foraminiferal compositions are also plotted using (a) the simple ZADR model  and (b) the ZADR model. The black lines refer to Triloba, the red lines to Obesa, the green lines to Pachyderma and the blue lines to Atlantica.}
\label{fig2}
\end{figure}

In Figure \ref{fig2}, the fitted foraminiferal compositions are plotted using the simple ZADR model and the ZADR model.  We have chosen the first component (by default) to be the reference component for all the other components in the ZADR model. Thus, the interpretation of the parameters (see Table \ref{tab1}) of the components is with respect to the first component. From the column of the constants we can understand that when the water depth is 1 metre and thus the logarithm of 1 is 0, Triloba (reference component) has the highest composition whereas Pachyderma has the lowest. As for the slopes, we can say that Obesa and Pachyderma increase their percentages more than Triloba as the water depth increases, whereas Atlantica reduces its percentage in relation to Triloba. Finally, when the ZADR (\ref{mixdirireg3}) is employed, the fitted $\phi_js$ decrease as the logarithm of the water depth increases as seen in Table 4.

The log-likelihood for the simple ZADR and ZADR models are 124.040 and 125.877 respectively, indicating no significant gain of the more complex model over the simpler model when compared with a $\chi^2$ with 1 degree of freedom. 

\subsubsection{Comparison of simple ZADR with other regression approaches}
Table \ref{tab2} shows the previously computed regression estimates of the simple ZADR model and of the Aitchison regression model applied to the zero value imputed data. We can see that the estimated coefficients are not that different to the ones obtained by the simple ZADR model. The coefficients of the Kent regression \citet{scealy2011} are not directly comparable to these ones and that is why we have not presented them here. The reason being is that \citet{scealy2011} did not use the additive log-ratio transformation (\ref{regalpha}) but rather the multiplicative log-ratio transformation. We can use their fitted values though for model fitting comparison.   

\begin{table}[!ht]
\caption{ \label{tab2} Parameter estimates for the the simple ZADR model and of the Aitchison regression model applied to the zero value imputed data. The standard error appears inside the parentheses.} 
\centering
\begin{tabular}{c|cc|cc|cc} \hline\hline
& \multicolumn{2}{c}{Simple ZADR model} & \multicolumn{2}{c}{Aitchison model}   \\  \hline                    
Response variable  & Constant   & Slope      & Constant       & Slope  \\  \hline  
Obesa      & -1.225(0.348)  & 0.117(0.131)   & -1.319(0.369)  & 0.121(0.140)    \\ [5pt]
Pachyderma & -2.392(0.463)	& 0.087(0.173)   & -3.065(0.755)  & 0.083(0.288)    \\ [5pt] 
Atlantica  & -2.298(0.494)  & -0.046(0.193)  & -2.596(0.540)  & -0.208(0.206)   \\ [5pt]   \hline \hline
\end{tabular}
\end{table}
 
Table \ref{tab2b} contains the Kulback-Leibler divergence and the $L_2$ norm  between the observed and the fitted compositional data using the simple ZADR model, the Kent model suggested by \citet{scealy2011} and the Aitchison model  applied to the data after the parametric zero value imputation. 

\begin{table}[!ht]
\caption{ \label{tab2b} Kulback-Leibler divergence and the $L_2$ norm between the observed and the fitted compositional data.} 
\centering
\begin{tabular}{c|ccccc} \hline\hline
Model         & Kulback-Leibler divergence & $L_2$ norm         \\       \hline                    
Simple ZADR   & 3.126                      & 1.065  \\ [5pt]
Kent          & 3.172	                   & 1.062  \\ [5pt] 
Aitchison     & 3.556                      & 1.173 \\ [5pt] \hline \hline
\end{tabular}
\end{table}

\begin{figure}[!ht]
\centering
\begin{tabular}{cc}
\includegraphics[scale=0.5,trim=0 20 20 20]{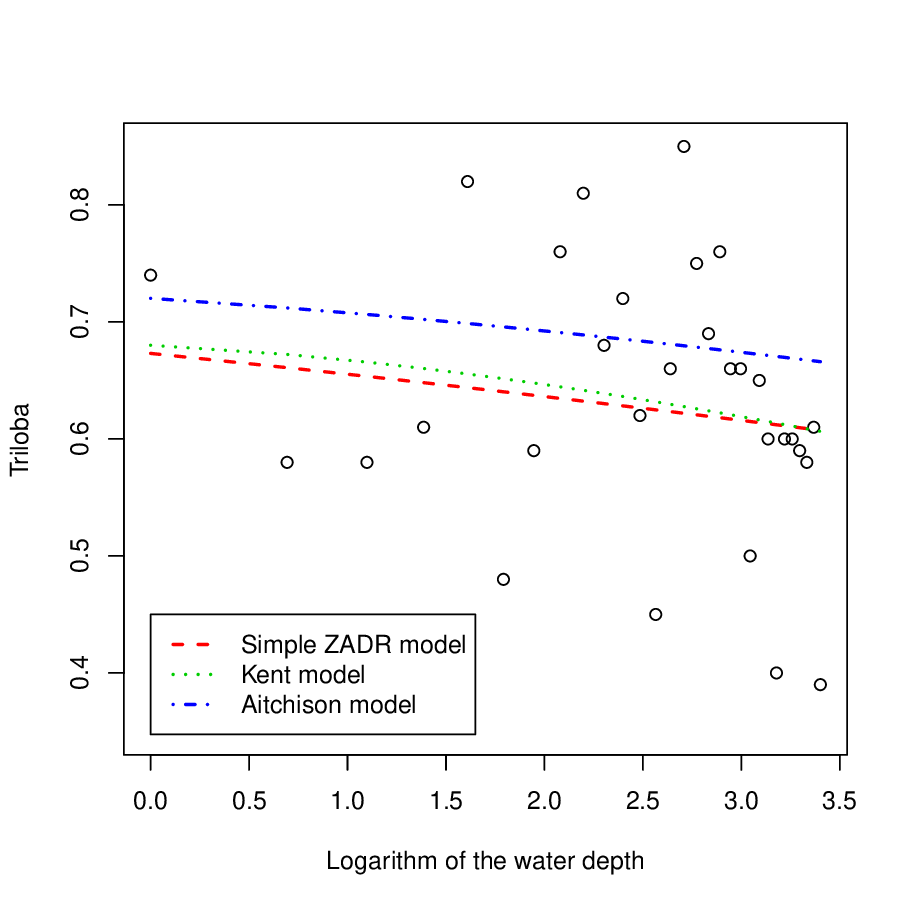} &
\includegraphics[scale=0.5,trim=0 20 20 20]{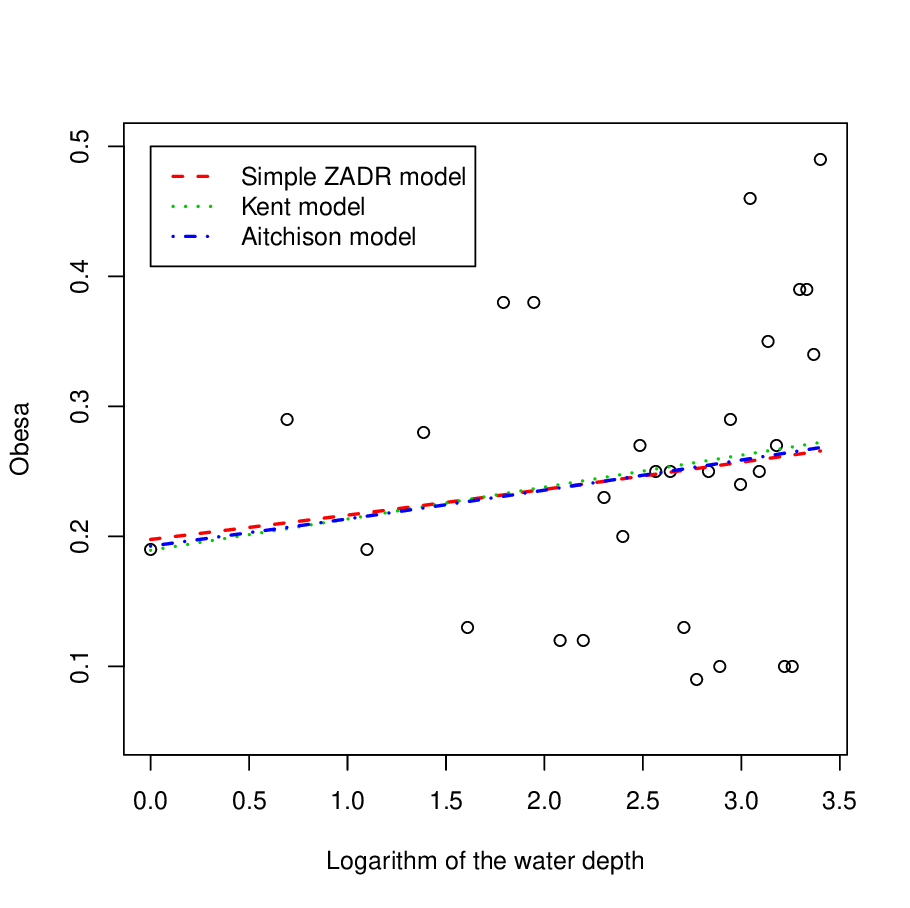}   \\
\footnotesize{(a)}   &  \footnotesize{(b)}               \\   
\includegraphics[scale=0.5,trim=0 20 20 20]{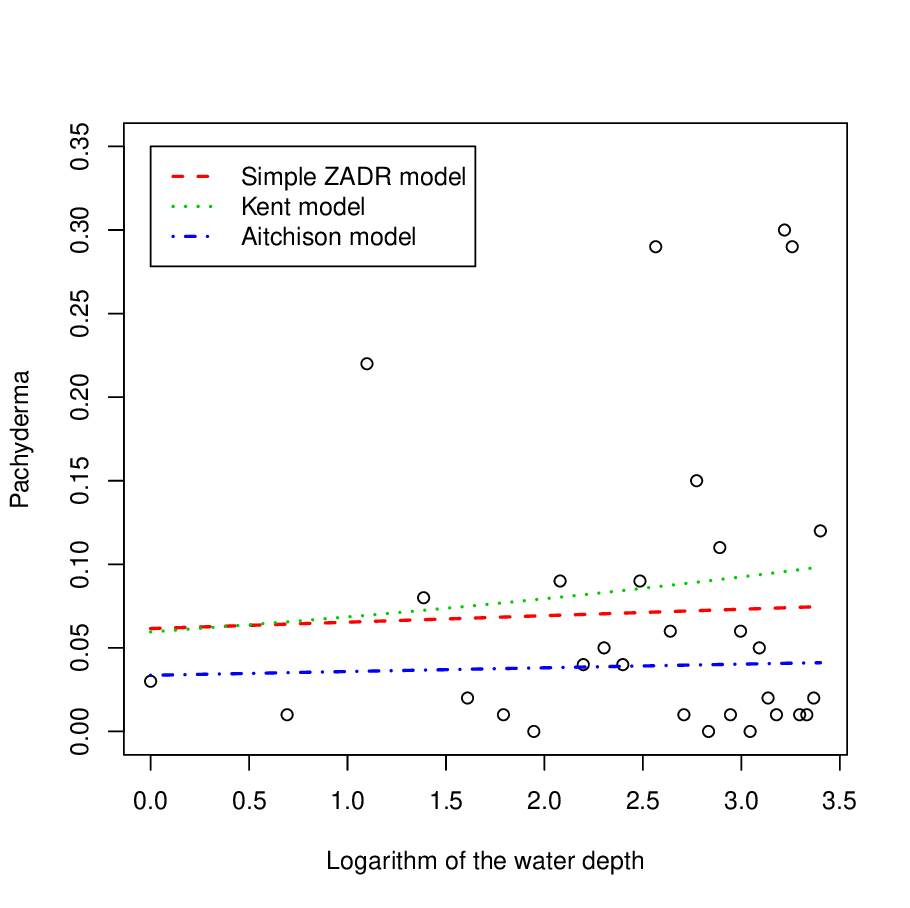} &
\includegraphics[scale=0.5,trim=0 20 20 20]{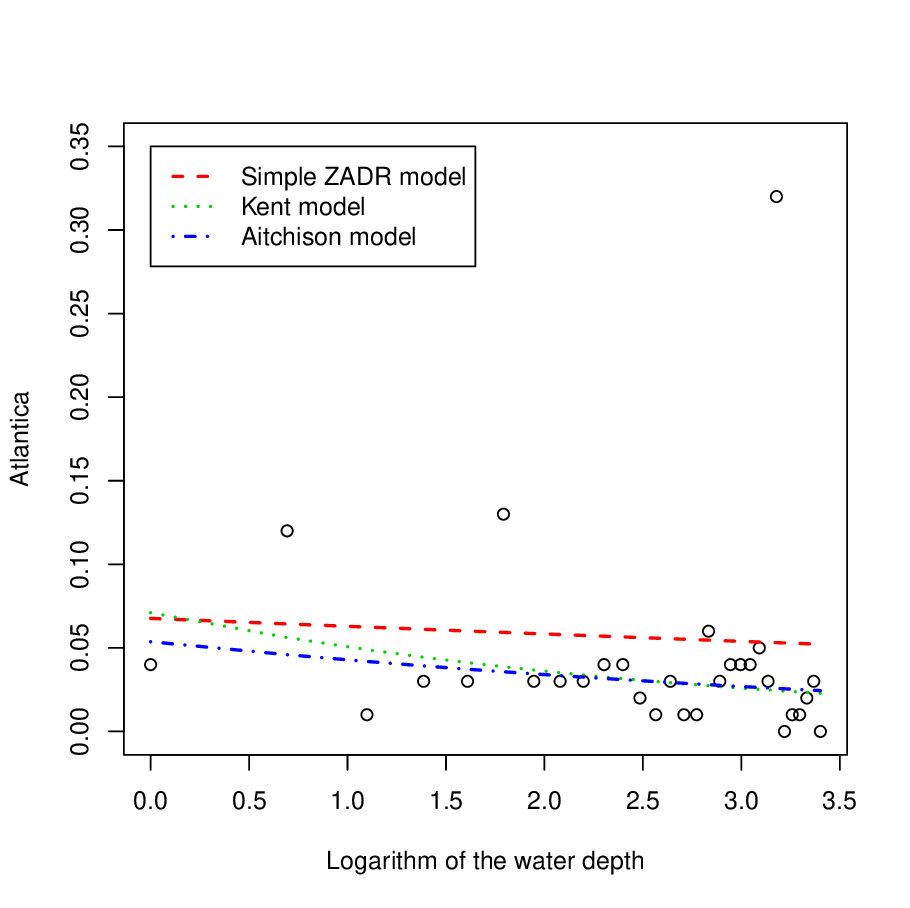}   \\
\footnotesize{(c)}   &  \footnotesize{(d)}               \\  
\end{tabular}
\caption{Fitted foraminiferal compositions of the simple ZADR, the Kent and the Aitchison model. Each component of the data appears in each graph.}
\label{fig3}
\end{figure}

\subsection{Real data analysis 4: Arctic lake data}
In this final real data analysis example we will use the Arctic lake data from \cite[pg.~359]{ait2003}. This example is a graphics based one. We want to show graphically, the "break down" of the Aitchison model when there are many zeros in the data, in contrast to the two ZADR models and the Kent model \citep{scealy2011}.

\begin{figure}[!ht]
\centering
\begin{tabular}{cc}
\includegraphics[scale=0.5,trim=40 40 50 40]{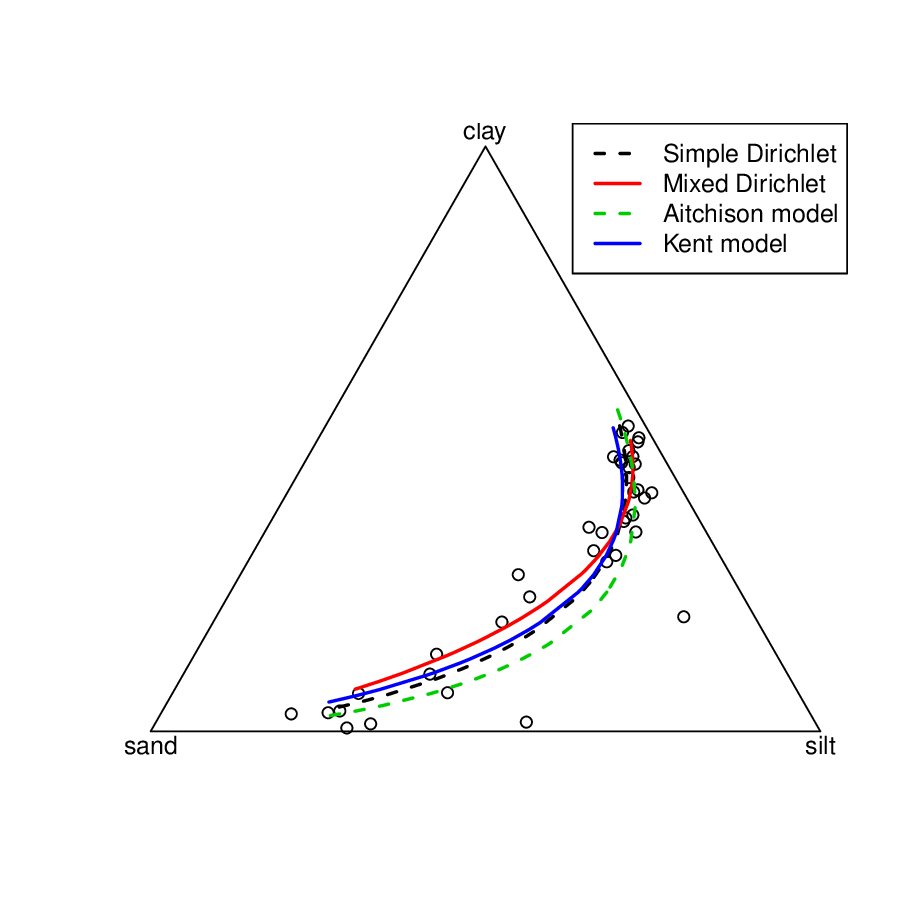} &
\includegraphics[scale=0.5,trim=40 40 50 40]{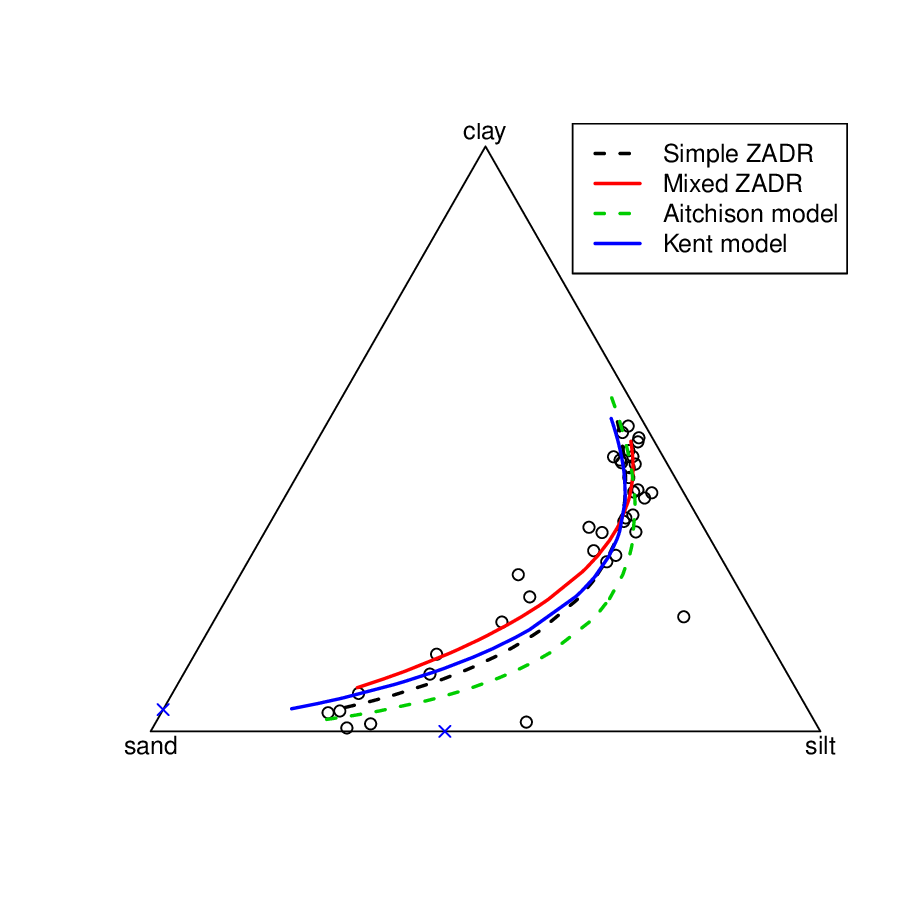}  \\
\footnotesize{(a)}   &  \footnotesize{(b)}              \\
\multicolumn{2}{c}{\includegraphics[scale=0.5,trim=40 40 50 40]{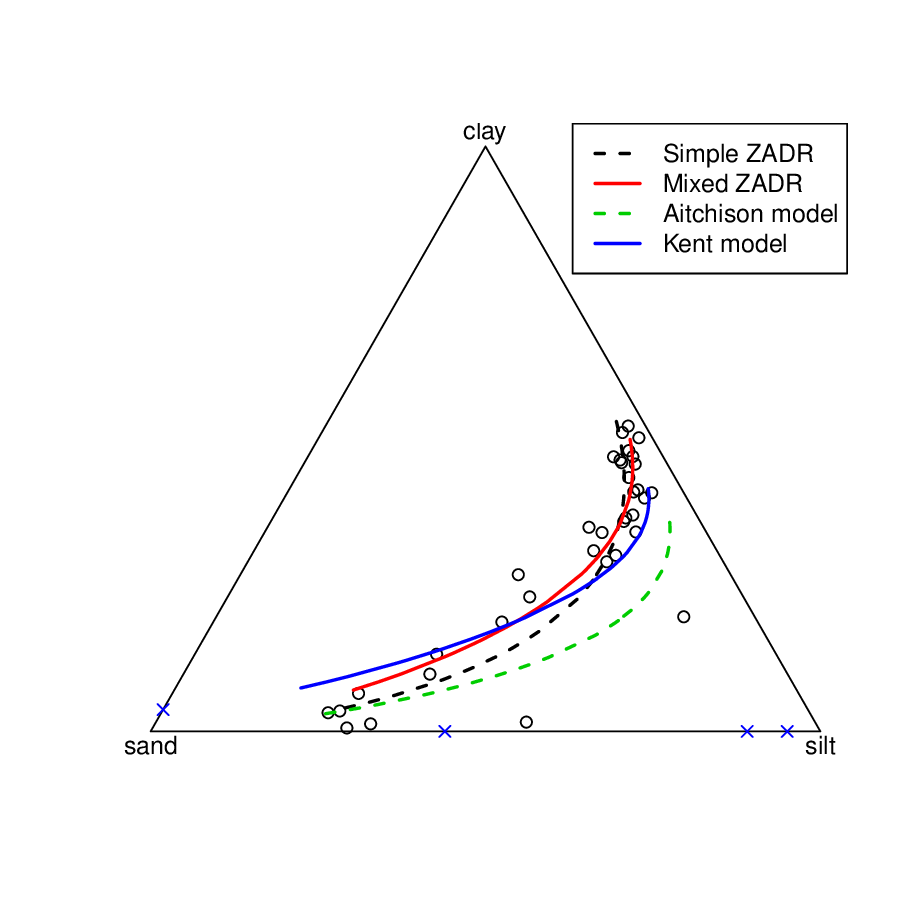} }   \\     
\multicolumn{2}{c}{\footnotesize{(c)} }   
\end{tabular}
\caption{Fitted Arctic lake compositions of the simple ZADR and ZADR, the Kent and the Aitchison model. As we move towards the depth of the lake, the data go from high values of sand to low values of sand and the percentages are allocated mostly between the silt and clay components. 
The first graph (a) shows the data as they are, without zeros. The Dirichlet models are the classical regression models. The second graph (b) shows the data with two imposed zeros and the third graph (c) shows the data with 4 imposed zeros.}
\label{fig4}
\end{figure}

The composition of an Arctic lake in sand, silt and clay is recorded at samples taken at 39 different water depths. Close to the surface, the percentage of sand is high and as we move deeper, this percentage reduces and the silt and clay compositions become relevantly similar. To see this graphically, one may choose any of the three ternary plots in Figure (\ref{fig4}) and see the lines from left to right. 

Four regression models (simple ZADR, ZADR, Aitchison's model and the Kent model) were fit to the original data which contained no zeros. Two and then 4 zeros were imposed and the models were fit again.  Figure \ref{fig4} shows the results of the 4 models not adjusted for zeros (Figure \ref{fig4}(a)) and adjusted for zeros (Figures \ref{fig4}(b) and \ref{fig4}(c)). What is most evident is that the Aitchison model (Figure \ref{fig4}(a)) is affected by the two points at the bottom right. In Figure \ref{fig4}(c) we can see that this regression line becomes more curved towards these two points. Finally, in the last ternary plot (Figure \ref{fig4}(c)) the regression line of the Aitchison model has lost the bulk of the data, since the extra two zeros were imposed to points with low sand percentage. This means that these two points appear to be highly influential for this model. The ZADR models though, seem not to have been affected at all. 

\subsection{Simulation Study: A comparison of simple ZADR and 
Aitchison regression}
This small scale simulation study focuses on the fit of the simple ZADR model versus the fit of the Aitchison regression model. The fit is measured using the Kullback-Leibler divergence of the observed data from the fitted data (see Equation (\ref{KL})). 

For a given number of components and sample sizes, we first generated data from a $\text{Dir}\left(\phi, \mathbf{a}^* \right)$ distribution where $\mathbf{a}^*$ are defined in (\ref{regalpha}) and $\phi=10$ is fixed throughout. We secondly generated data from a multivariate normal distribution, $MVN_{D-1}\left(\pmb{\mu},\pmb{\Sigma} \right)$, where $\pmb{\mu}={\bf BX}$ and the inverse of the additive log-ratio transformation (\ref{alr}) is used to map the data into the simplex.  In both cases the design matrix contained two predictor variables (that is,  ${\bf X}=\left(1,{\bf x}_1,{\bf x}_2\right)$) whose values were randomly generated from two normal distributions. The ${\bf B}$ matrix of parameters were also generated from a normal distribution. Finally, in the multivariate normal distribution, the covariance matrix $\pmb{\Sigma}$ is a diagonal matrix with elements coming from an exponential distribution (the choice of this distribution and of the diagonal structure was decided at random and with no purpose, or any motivation behind it).   

We examined six sample sizes, namely $n=\left(30,50,100,200,350,500\right)$ and three different composition lengths,\ $D=\left(3,5, 10\right)$. We also added differing amounts of zeros in the generated data by assigning $\pi=\left( 10\%,20\%,30\%,40\%,50\% \right)$ values a zero, and then scaling observations that were assigned a zero so that they summed to 1.  For compositions chosen to contain a zero, when $D=3$, one component was chosen, when $D=5$, two components were chosen at random and when $D=10$, five components were chosen at random. Hence, we considered 6 sample sizes, 3 composition lengths and 5 different percentage levels of zero values. For each of these combinations ($6 \times 3 \times 5$), $1000$ simulations were performed and each time the simple ZADR and Aitchison regression models were fitted. The proportion of times the Kullback-Leibler divergence of the ZADR model was less than the Kullback-Leibler divergence of the Aitchison model is reported.   

\begin{figure}[!ht]
\centering
\begin{tabular}{ccc}
\multicolumn{3}{c}{Logistic normal distribution simulated data }  \\     
\includegraphics[scale=0.3,trim=0 0 0 0]{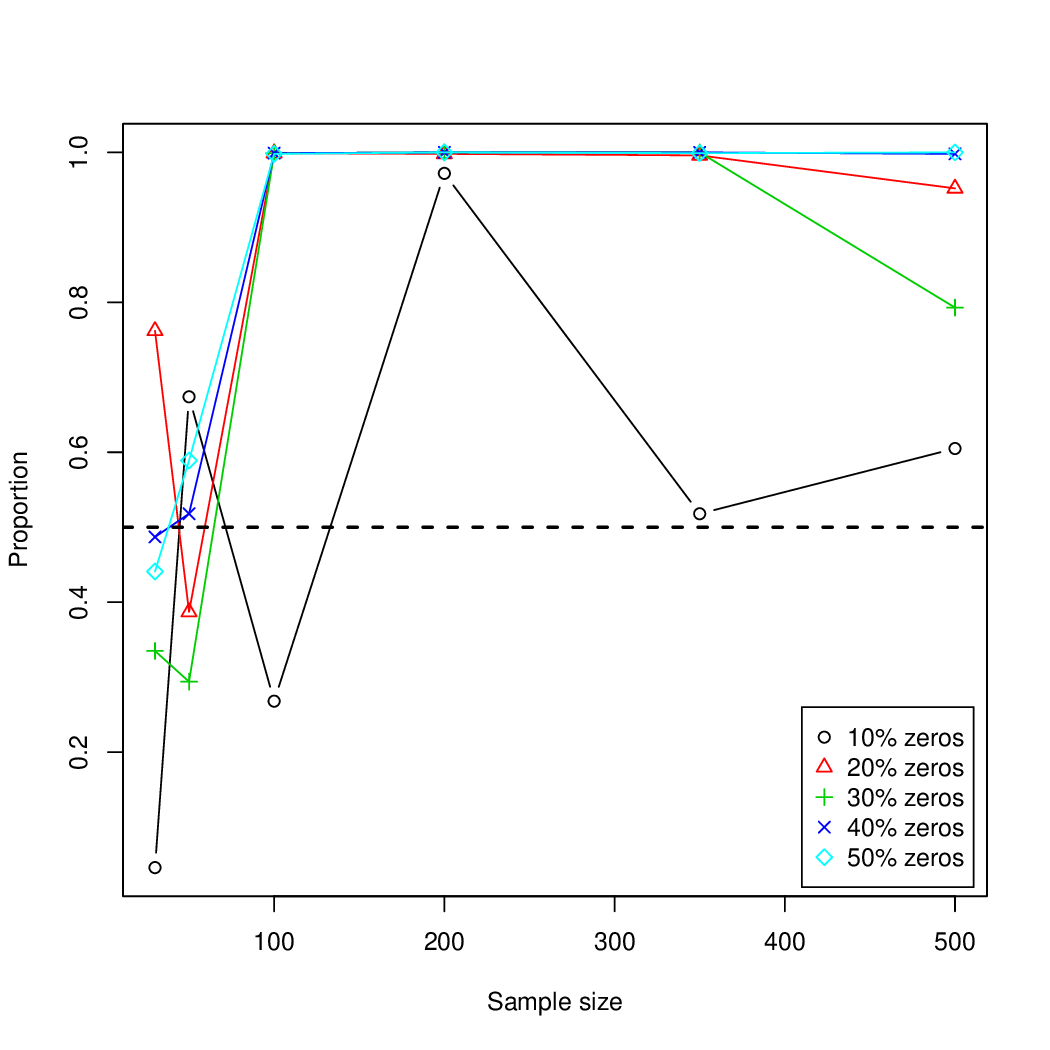} &
\includegraphics[scale=0.3,trim=0 0 0 0]{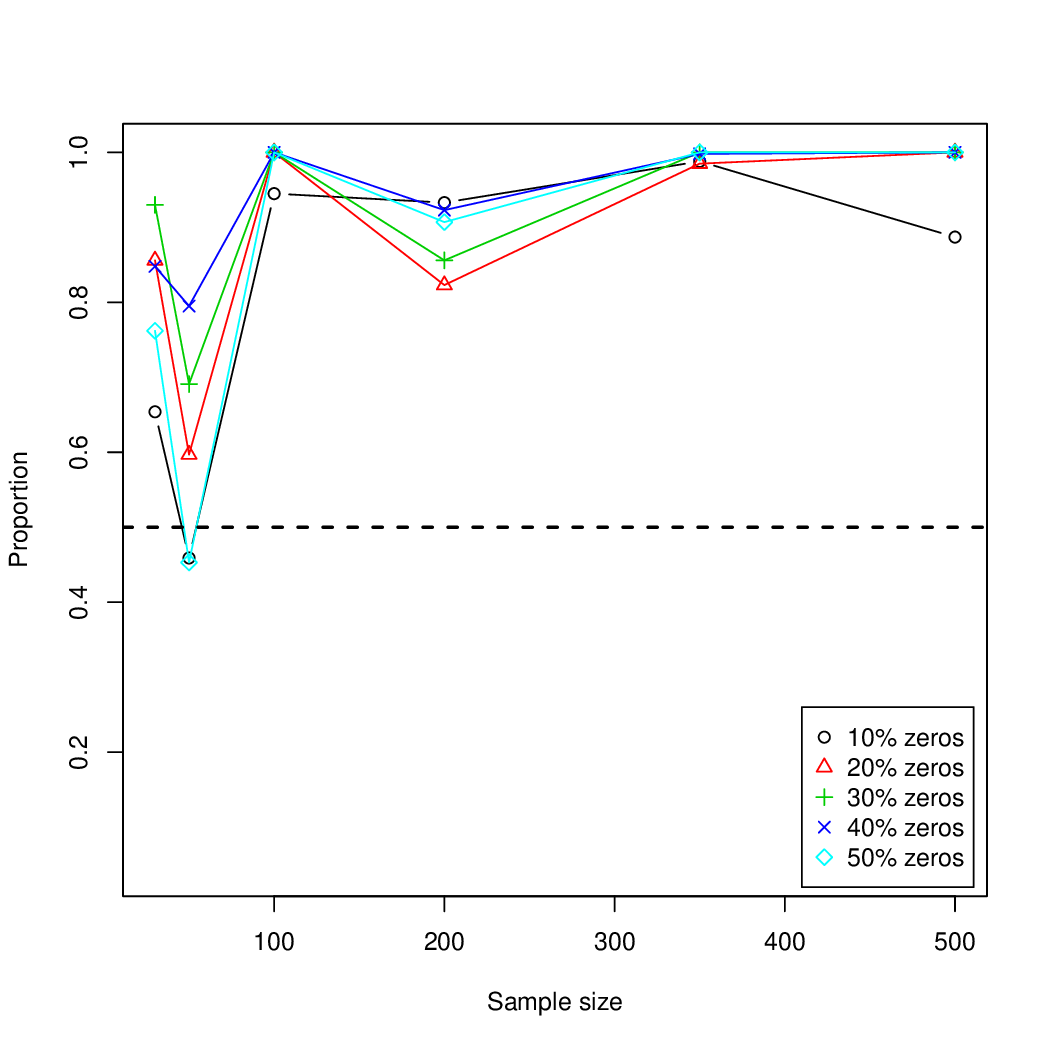} &             
\includegraphics[scale=0.3,trim=0 0 0 0]{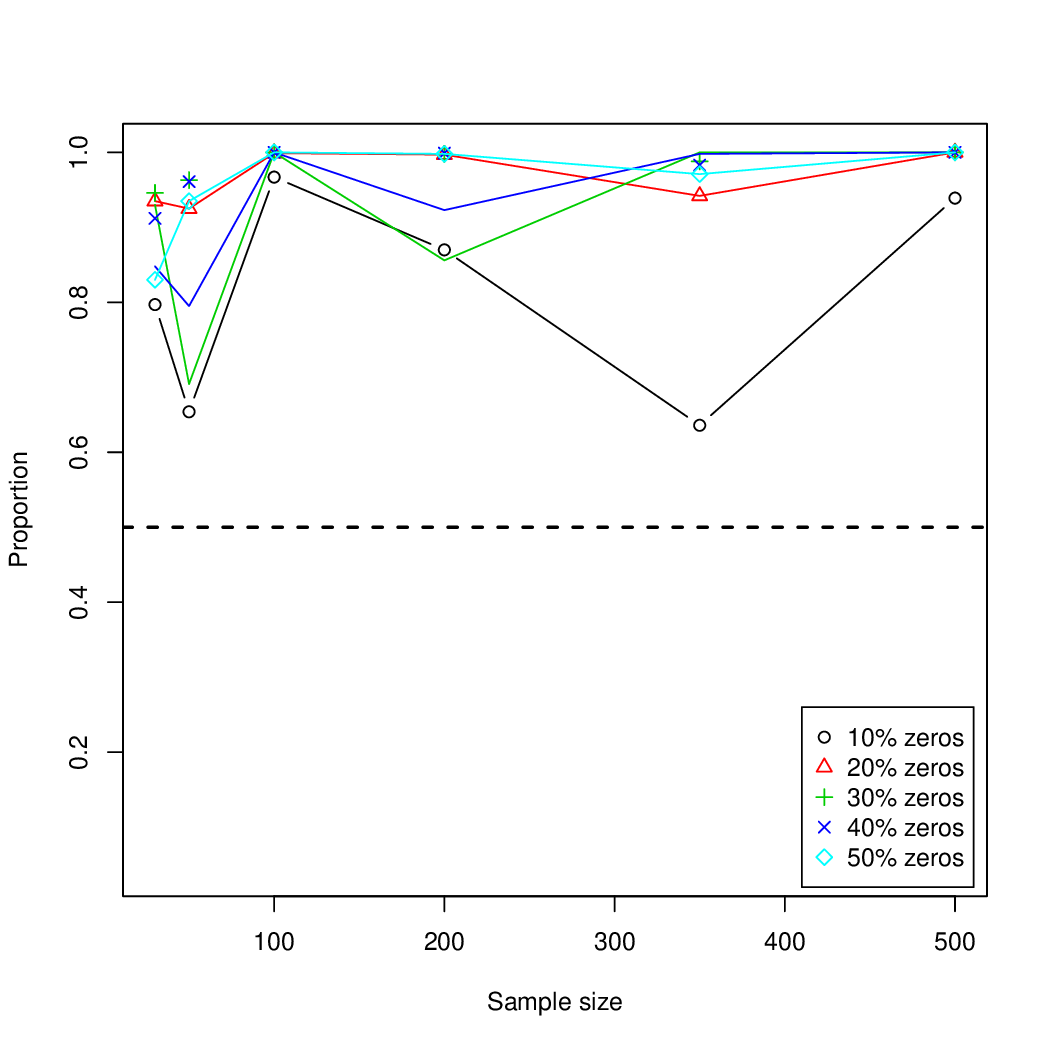}               \\
\footnotesize{3 Components}   &  \footnotesize{5 Components}  &  \footnotesize{10 Components}     \\
\multicolumn{3}{c}{Dirichlet distribution simulated data }        \\     
\includegraphics[scale=0.35,trim=0 0 0 0]{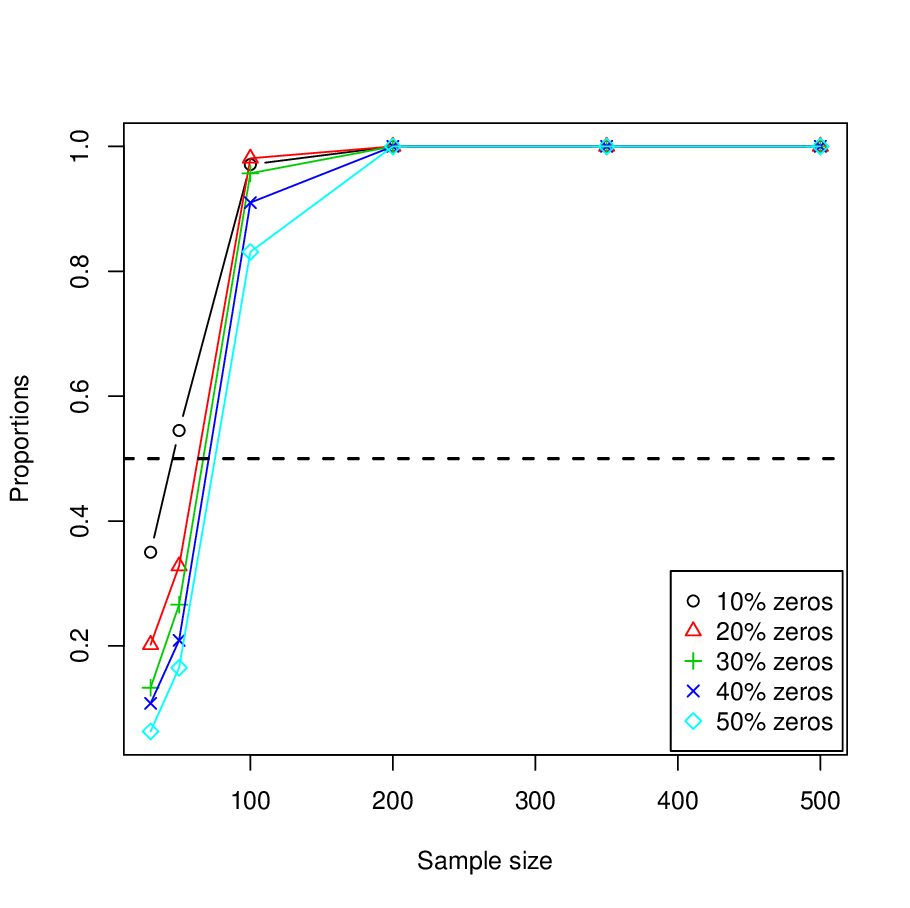} &   &
\includegraphics[scale=0.35,trim=0 0 0 0]{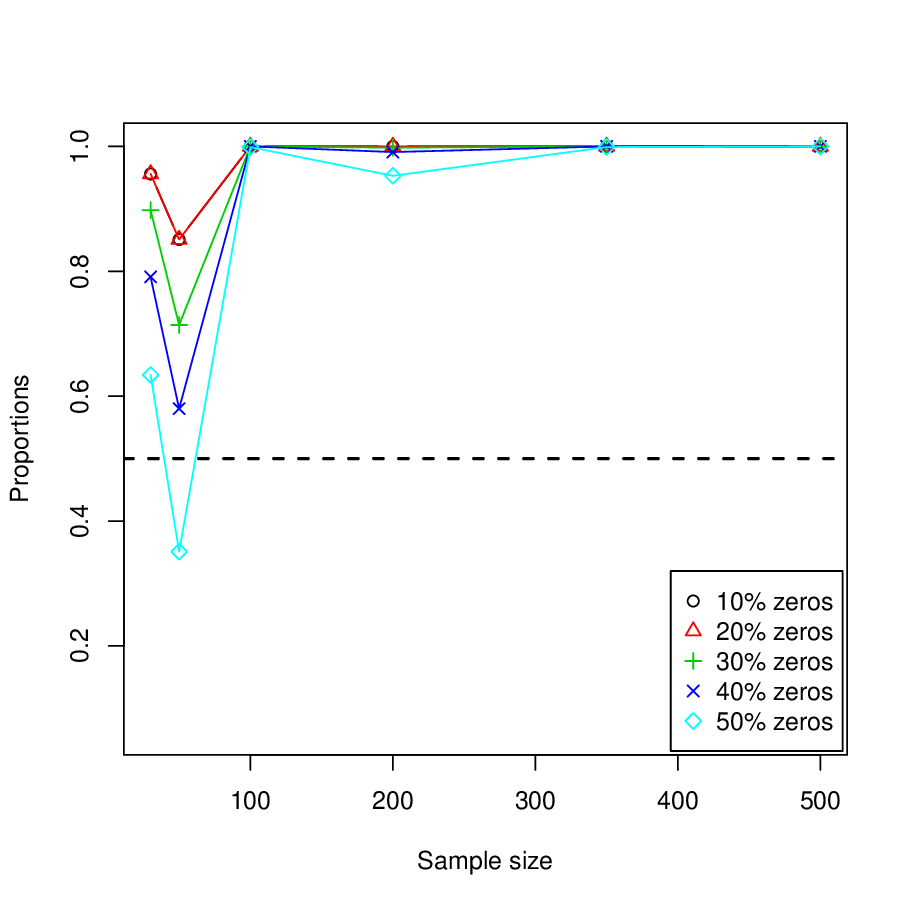}               \\
\footnotesize{3 Components}   &   &  \footnotesize{5 Components}  \\
\end{tabular}
\caption{Proportion of times the Kullback-Leibler divergence of the ZADR model is less than the Kullback-Leibler divergence of the Aitchison model. The horizontal line is drawn at 0.5. When there are $10$ components in the Dirichlet simulated data case, the proportion was $1$ for all sample sizes.}
\label{fig5}
\end{figure}

The results are presented in Figure \ref{fig5}. For the Dirichlet simulated data we see that as the sample size increases, the proportion of times the Kullback-Leibler divergence of the ZADR model is less than the corresponding divergence of the Aitchison model increases and reaches 1. This is also the case for the logistic normally simulated data when there are 5 components. When there are 3 components, the proportion of times ZADR does better reaches 1 in the case of many zeros ($40\%$ and $50\%$). 

A nice property of the Dirichlet distribution is that if some of its components are assigned a zero, the other non zero components follow a Dirichlet distribution with the same parameters. This is exactly what is used here and why it works very well in the case of Dirichlet simulated data. In the case of the logistic normal, we do not have this result.  However, when there are a lot of zeros, we believe that the zero replacement technique \cite{templ2011} may not work very well and that is why ZADR does better. 

\section{Discussion} 
The Dirichlet distribution offers a nice way of handling zeros and allows us to perform regression without having to substitute the zero values with small quantities. The goal is to treat the zero values without changing their values and to some extent the values of the other components in the same compositional vector. This approach is particularly valuable in applications where zeros are truly zero values.

A key point worthy of mention is the assumption of common estimates (both the concentration and the set of beta coefficients) across all patterns of zeros and non zeros. With $D$ components there are $\sum_{i=1}^{D-2}\binom {D}{i}$ possible zero patterns. It might seem a rather restrictive assumption, but in fact it is a rather necessary one since otherwise, for one, there would need to be a large number of observations with each pattern of zeros. Patterns with only one or two observations could not be handled. Secondly, as the number of different patterns increased, the number of parameters needed to be estimated would grow quickly. For similar reasons, it is not practical to link covariates to the estimates of the patterns of zeros. In our examples and data we have come across, it is usually the case that some patterns of zero values are repeated. On the other hand it is not necessary either that a large number of observations are zero free. 

A known drawback of the Dirichlet distribution and thus of regression is that the covariance matrix allows only negative correlations. This restriction is also met in the multinomial logistic regression.  While in some situations this constraint may be impractical, there are also disadvantages associated with the other current models. For example, as previously pointed out, the log ratio approach in \citet{bear2015} requires that one component is never zero while the model in \citet{scealy2011} is more difficult to fit and interpret. A more serious drawback is that when no covariates are present the Dirichlet distribution cannot capture the curvature in the simplex \citep[pg.~59]{ait2003}.  More specifically, when a line segment in $\mathbb{R}^2$ is plotted in $\mathbb{S}^2$ it becomes a curve. Unlike the logistic normal, Dirichlet can handle only linear relationships within the simplex. This is not the case, however, in our situation where covariates are present and the multivariate logit link function (\ref{regalpha}) is used. While \cite{ait2003} has been quite critical of the Dirichlet distribution, \cite{campbell1987} has shown that in the regression setting the Dirichlet distribution works sufficiently well. Apart from the results we demonstrated, we have seen this empirically.  See also \cite{maier2014} and \cite{gueorguieva2008} for additional examples.

As appealing the Dirichlet regression might seem, we have to note that the Dirichlet distribution is not a member of the linear exponential family \citep{gourieroux1984}.  Hence any regression model based on a distribution not satisfying this property is not robust to distribution miss-specification \citep{murteira2016}. However, under correct specification, the asymptotic distribution of the estimated parameters is a multivariate normal distribution whose covariance matrix is the inverse of the Fisher's information matrix \citep{murteira2016}. This means that in this case only the estimates are consistent. The same holds true in our case since we maximise the Dirichlet log-likelihood in order to obtain the regression parameters, but we need an extra assumption to ensure the property of consistency. That is, the compositional vectors which contain zero values, and are thus modelled by a conditional Dirichlet with less dimensions, come from a Dirichlet with the same parameters as the full Dirichlet. We have not proved that the conditions for this assumption to hold are true. 

The inflated beta distribution \citep{ospina2010} is not related to the conditional Dirichlet distribution we describe here. In the inflated beta distribution (with or without covariates) zero values are allowed and the same is true in the logistic regression. But when we move to higher dimensions, the Dirichlet distribution does not allow a vector of zeros or a vector of $1s$. The only generalisation that can happen is the $1$-inflated Dirichlet distribution, that is, when there is a compositional vector having the value of $1$ in one of its components and $0$ elsewhere. \citet{butler2008} analyse a compositional dataset with only $3$ components and they observe this case. However, particularly when we move to higher dimensions, it seems unlikely that this situation would arise in practice and, even in this case, we could simply put a point mass distribution on $1$ and ignore the zero values of the other components just like we did here. 

While a Bayesian approach is beyond the scope of this model, it could offer a richer inference at the expense of a much harder to fit model.  

\section{Conclusions}
In this paper we suggested a zero adjusted Dirichlet regression (ZADR) model for modelling compositional data with covariates when zero values are present. The importance of this simple approach is important, since no modification of the data is necessary, such as zero value replacement (or imputation). This means that no extra variance is introduced and most importantly the observed data are not at all changed or distorted to the slightest. 

The extension of this Dirichlet regression involved linking the precision parameter ($\phi$) to the same covariates, thus leading to the so called ZADR model as opposed to the simple ZADR model. 

Examples with real data illustrated the applicability of the two models and the interpretation of their coefficients. For the first two examples, the Aitchison model performed slightly better than the simple ZADR model. For the next two examples though, the simple ZADR did either slightly or much better. A comparison with other methods showed that the results were similar to the simple ZADR model. As for the simulation study where the distribution of the generated data was distorted by the inclusion of random zero values, the simple ZADR model performed better the majority of the time. 

The question though remains, should we treat zeros as rounded and try to impute their missing value or should we treat them naturally? We believe the latter option is preferable and one reason for this relates to the compositional nature of the data since changing the value of one component results in changes in all the other components. Our proposed approach is not a difficult one and seems to work adequately.

\bibliographystyle{apa}

\end{document}